\documentclass[12pt,a4paper]{article}
\usepackage{t1enc}
\usepackage{amscd}
\usepackage{amsmath,amssymb}
\usepackage{theorem}
\usepackage{graphicx,url}
\usepackage{subfig}

\renewcommand{\v}[1]{\mathbf{#1}}
\renewcommand{\emph}{\textbf}

\theorembodyfont{\upshape}
\newtheorem{Thm}{Theorem}
\newtheorem{Def}{Definition}
\newtheorem{Exa}{Example}
\newtheorem{Rem}{Remark}
\newtheorem{Lem}{Lemma}

\newtheorem{Sta}{Statement}
\newcommand{\Proof}[1]{{\textit{Proof. }#1}}

\newcommand{\torol}[1]{}

\newcommand{\revreak}[4]{#1{\ }{\overset{#2}{\underset{#3}{\rightleftharpoons}}#4}}

\newcommand{\eb}{\mathbf{e}}
\newcommand{\N}{\mathbb{N}}
\newcommand{\R}{\mathbb{R}}
\newcommand{\In}[1]{\textbf{\tt{In}[#1]:= }}
\newcommand{\Out}[1]{\tt{Out}[#1]= }
\newcommand{\mtt}[1]{\mbox{\textbf{\tt{#1}}}}
\newcommand{\ttt}[1]{\textbf{\tt{#1}}}

\title{Microscopic Reversibility or Detailed Balance\\
in Ion Channel Models}
\author{
Ilona Nagy\\
Department of Mathematical Analysis\\
Budapest University of Technology and Economics,\\
Budapest, Egry J. u. 1., HUNGARY, H-1111\\
\url{nagyi@math.bme.hu} and\\
J\'anos T\'oth\\
Department of Mathematical Analysis\\
Budapest University of Technology and Economics,\\
Budapest, Egry J. u. 1., HUNGARY, H-1111 and\\
Laboratory for Chemical Kinetics
Eötvös University\\
Budapest, Pázmány Péter sétány 1/A, HUNGARY, H-1117\\
\url{jtoth@math.bme.hu}}
\date{}

\begin{document}
{\LARGE
\noindent \textbf{Title Page}

\noindent \textbf{Title:}
Microscopic Reversibility or Detailed Balance
in Ion Channel Models

\noindent \textbf{Authors:}

Ilona Nagy\footnote{(author designated to review proofs)
Department of Mathematical Analysis,
Budapest University of Technology and Economics,
Budapest, Egry J. u. 1., HUNGARY, H-1111,
Phone: 361 463-5141,
Fax: 361 463 3172,
E-mail:
\url{nagyi@math.bme.hu}}

Tóth, J.\footnote{
Department of Analysis, Budapest University of Technology and Economics,
Egry J. u. 1., Budapest, Hungary, H-1111 and
Laboratory for Chemical Kinetics, Eötvös Loránd University,
Pázmány P. sétány 1/A, Budapest, Hungary, H-1117}

\noindent \textbf{Running Head}: Detailed balance in ion channels}
\vfill\eject

\maketitle

\begin{abstract}
Mass action type deterministic kinetic models of ion channels are
usually constructed in such a way as to obey the principle of
detailed balance (or, microscopic reversibility) for two reasons:
first, the authors aspire to have models harmonizing with
thermodynamics, second, the conditions to ensure detailed balance
reduce the number of reaction rate coefficients to be measured. We
investigate a series of ion channel models which are asserted to
obey detailed balance, however, these models violate mass
conservation and in their case only the necessary conditions (the
so-called circuit conditions) are taken into account. We show that
ion channel models have a very specific structure which makes the
consequences true in spite of the imprecise arguments. First, we
transform the models into mass conserving ones, second, we show that
the full set of conditions ensuring detailed balance (formulated by
Feinberg) leads to the same relations for the reaction rate
constants in these special cases, both for the original models and
the transformed ones.
\end{abstract}
\tableofcontents\eject
\section{Introduction}
\subsection{Detailed balancing or microscopic reversibility}

At the beginning of the 20th century it was Wegscheider
\cite{wegscheider} who gave the formal kinetic example \(
\revreak{A}{k_1}{k_{-1}}{B}\quad \revreak{2A}{k_2}{k_{-2}}{A+B} \)
to show that in some cases the existence of a positive stationary
state alone does not imply the equality of all the individual
forward and backward reaction rates in equilibrium: a relation
($\frac{k_1}{k_{-1}}=\frac{k_2}{k_{-2}}$) should hold between the
reaction rate coefficients to ensure this. Equalities of this kind
will be called (and later exactly defined) as \emph{spanning forest
conditions} below. Let us emphasize that violation of this equality
does not exclude the existence of a positive stationary state; it
exists and it is unique for all values of the reaction rate
coefficients, see the details in subsection \ref{db}.
\begin{figure}[h!]
  \centering
  \subfloat[kjhkjh][The Wegscheider reaction]{\label{fig:wegscheider}\includegraphics[width=0.4\textwidth]{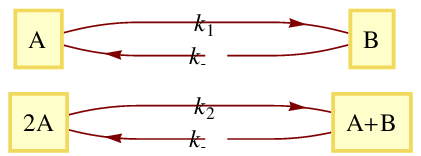}}
  \subfloat[][The triangle reaction]{\label{fig:triangle}\includegraphics[width=0.4\textwidth]{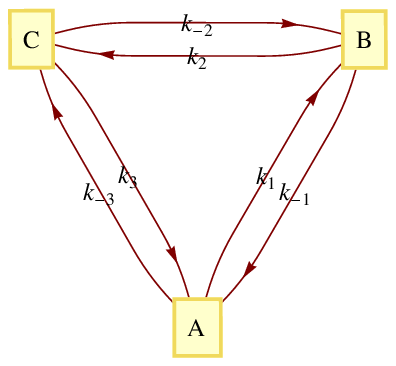}}
  \caption{}
  \label{fig:1}
\end{figure}

A similar statement holds for the reversible triangle reaction in
Fig. \ref{fig:triangle}.
The necessary and sufficient condition for
the existence of such a positive stationary state for which all the
reaction steps have the same rate in the forward and backward
direction is $k_1k_2k_3=k_{-1}k_{-2}k_{-3}.$
Equalities of this kind
will be called (and later exactly defined) as \emph{circuit
conditions} below.
Again, violation of this equality does not
exclude the existence of a positive stationary state; it exists and
is unique for all values of the reaction rate coefficients, see
the details in subsection \ref{db}.

A quarter of a century after Wegscheider the authors Fowler and
Milne \cite{fowler} formulated in a very vague form a general
principle called the \emph{principle of detailed balance} stating
that in real thermodynamic equilibrium all the subprocesses
(whatever they mean) should be in dynamic equilibrium separately in
such a way that they do not stop but they proceed with the same
velocity in both directions. Obviously, this also means that time is
reversible at equilibrium, that is why this property may also be
called \emph{microscopic reversibility}.

A relatively complete
summary of the early developments was given by Tolman
\cite{tolman}.

The modern formulation of the principle accepted by IUPAC \cite{goldloeningmacnaughtshemi}
essentially means the same: ``The principle of microscopic reversibility at equilibrium states that, in a system at equilibrium,
any molecular process and the reverse of that process occur, on the average, at the same rate.''

Neither the above document nor the present authors assert that the principle should hold without any further assumptions;
for us it is an important hypothesis the fulfilment of which should be checked individually in different models.

It turned out that in the case of chemical reactions this general
principle can only hold if both the spanning tree conditions and the
circuit conditions are fulfilled.
However, it became a general belief among people dealing with reaction kinetics 
that the circuit conditions alone are not only necessary but also
sufficient for all kinds of reactions: Wegscheider's example proving
the contrary was not known well enough.
Vlad and Ross \cite{vladross} draw the conclusions from the Wegscheider example in full generality,
but it was Feinberg \cite{feinberg} who gave the definitive solution of the problem in
the area of formal kinetics: he clearly formulated, proved and
applied the two easy-to-deal-with sets of conditions which together
make up a necessary and sufficient condition of detailed balance
(for the case of mass action kinetics). In other words, he completed
the known necessary condition (the circuit conditions) with another
condition (the spanning forest conditions) making this sufficient,
as well.

The reason why the false belief is widespread is that
in case of reactions with deficiency zero the circuit
conditions alone are also sufficient not only necessary,
and most textbook examples have deficiency zero.
\subsection{Ion channel models}
Recent papers on formal kinetic models of ion channel gating show
that people in this field think that the principle of detailed
balance or microscopic reversibility should hold. (However, some
authors do not consider the principle of microscopic reversibility
indispensable, e. g. Naundorf et al. \cite[Supplementary Notes 2,
Fig. 3SI(a), page 4]{naundorf} provides a channel model which is not
even reversible, let alone detailed balanced.) This may be supported
either by a \emph{theoretical} argument: they should obey the laws
of thermodynamics, or by a \emph{practical} one: if the principle
holds one should measure fewer reaction rate coefficients because one
also has the constraints implied by the principle. The second
argument seems to be the more important one in the papers by
Colquhoun et al. \cite{colquhoun2}, \cite{colquhoun1}.  However, the
principle is applied in an imprecise way: first, only the necessary
part consisting of the circuit conditions is applied, second, the
models are formulated in a way that they do not obey the principle
of \emph{mass conservation}. In the present paper we transform the
models into mass conserving ones, and apply the full set of
necessary and sufficient conditions. Our main result is that in
classes of models including all the known ion channel examples are
compartmental models, therefore they have zero deficiency at the
beginning, and being transformed into a mass conserving model they
have no circuits, therefore one has only to test the spanning forest
conditions. It is not less interesting that the spanning forest
conditions obtained for the transformed models are literally the
same as the circuit conditions for the original models.
\subsection{Stochastic models}
So far we had in mind only deterministic models (surely not speaking
of the general but vague formulation of Fowler and Milne). Turning
to stochastic models one possible approach is to check the
fulfilment of microscopic reversibility in the following way. Let us
suppose we have some measurements on a process, and present the data
with reversed time, finally use a statistical test to see if there
is any difference. This is an absolutely correct approach and has
also been used in the field of channel modeling \cite{rothberg}.
\subsection{Outline}
The structure of our paper is as follows. Section 2 gives a short
summary of the definitions used and presents Feinberg's theorem. In
Section 3 some usual ion channel models are transformed into
realistic models with mass conservation and with the help of a lemma
it is shown that \emph{in these special cases} the circuit
conditions for the origial systems and the spanning forest
conditions for the transformed systems lead to exactly the same
requirements. The question of the number of free parameters is also
discussed here. Finally an outlook and discussion follows in Section
4. The formal proof of our main result has been relegated to an Appendix.

Let us also mention that parts of our investigations has been
presented in a short, nonrigorous form in \cite{nagykovacstoth}.

\section{Tools to be used}
\subsection{Ion channels}
There is a difference in electric potential between the interior of
cells and the interstitial liquid. An essential part of the system
controlling the size of this potential difference is the system of
\emph{ion channels}: pores made up from proteins in the membranes
through which different ions may be transported via active and
passive transport thereby changing the potential difference in an
appropriate way. The models of these ion channels are usually
described in terms of formal reaction kinetics, thus we have to
present these notions first, then we shall be in the position to
present a few alternative models of ion channels.
\subsection{Basic definitions of formal kinetics}\label{subsec:frk}
Let us consider the reversible reaction
\begin{eqnarray}\label{reaction}
\sum_{m=1}^{M}\alpha(m,p)X(m)\rightleftharpoons\sum_{m=1}^{M}\beta(m,p)X(m)\quad
(p=1,2,\dots,P)
\end{eqnarray}
with $M\in\N$ chemical \emph{species}: $X(1),X(2),\dots,X(M);{\ }
P\in\N$ pairs of \emph{reaction steps}, $\alpha(m,p),\beta(m,p)\in\N_0{\ }
(m=1,2,\dots,M; p=1,2,\dots,P)$
\emph{stoichiometric coefficients} or \emph{molecularities}, and suppose its deterministic model
\begin{eqnarray}\label{ikde}
{c'}_m(t)&=&f_m(\v{c}(t)):=
\sum_{p=1}^P(\beta(m,p)-\alpha(m,p))(w_{+p}(\v{c}(t))-w_{-p}(\v{c}(t)))\\
c_m(0)&=&c_{m0}\in\R_0^+\quad(m=1,2,\dots,M)\label{ini}
\end{eqnarray}
---describing the time evolution of the concentration vs. time functions
$$
t\mapsto c_m(t):=[X(m)](t)
$$
of the species---is based on \emph{mass action type kinetics}:
\begin{eqnarray}
w_{+p}(\overline{\v{c}})&:=&k_{+p}\overline{\v{c}}^{\alpha(.,p)}:=
k_{+p}\prod_{\mu=1}^{M}\overline{\v{c}}_{\mu}^{\alpha(\mu,p)}\quad\\
w_{-p}(\overline{\v{c}})&:=&k_{-p}\overline{\v{c}}^{\beta(.,p)}:=
k_{-p}\prod_{\mu=1}^{M}\overline{\v{c}}_{\mu}^{\beta(\mu,p)}\\
(p&=&1,2,\dots,P).\nonumber
\end{eqnarray}
(\eqref{ikde} is also called the \emph{induced kinetic differential equation}
of the reaction \eqref{reaction}.)
The number of \emph{complexes} is the number of different
\emph{complex vectors} among $\alpha(.,p)$ and $\beta(.,p),$
i.e. it is the cardinality of the set
$$
\{\alpha(.,p);p=1,2,\dots,P\}\cup\{\beta(.,p);p=1,2,\dots,P\}
$$
and it is denoted by $N$. The \emph{Feinberg--Horn--Jackson graph}
(or, FHJ-graph, for short) of the reaction is obtained if one writes
down all the complex vectors (or simply the \emph{complexes}, the
formal linear combinations on both sides of \eqref{reaction})
exactly once and connects two complexes with an edge (or two
different edges pointing into opposite directions) if there is a
reaction step taking place between them. Let us denote the number of
connected components of this graph by $L.$

The \emph{stoichiometric space} is the linear subspace of
$\R^M$ generated by the \emph{reaction vectors}:
$\mathrm{span}\{\beta(.,p)-\alpha(.,p);p=1,2,\dots,P\};$
its dimension is denoted by $S.$ Finally, the nonnegative integer
$\delta:=N-L-S$ is the \emph{deficiency} of the reaction \eqref{reaction}.

Examples to show the meaning of the definitions follow.
\begin{Exa}[Simple bimolecular reaction]
In the simple reversible bimolecular reaction
$A+B\rightleftharpoons C$ we have
$M=3, P=1; X(1)=A, X(2)=B, X(3)=C;$ and
the complexes are $A+B$ and $C,$
thus the corresponding complex vectors are $(1,1,0)$ and
$(0,0,1).$
As $N=2, L=1, S=1;$ the deficiency of the reaction is 0.
\end{Exa}
\begin{Exa}[Triangle reaction]
In the triangle reaction (Fig. \ref{fig:triangle})
we have $M=3, P=3; X(1)=A, X(2)=B, X(3)=C;$ and
the complexes are $A,B$ and $C,$
thus the corresponding complex vectors are
$(1,0,0),(0,1,0)$ and $(0,0,1).$
As $N=3, L=1, S=2;$ the deficiency of the reaction is 0.
\end{Exa}
\begin{Exa}[Wegscheider]
In the Wegscheider reaction (Fig. \ref{fig:wegscheider}) we have
$M=2, P=2; X(1)=A, X(2)=B;$ and the complexes are $A,B,2A$ and
$A+B,$ thus the corresponding complex vectors are
$(1,0),(0,1),(2,0)$ and $(1,1);$ therefore the reaction vectors are
$(1,-1)$ and $(-1,1).$ As $N=4, L=2, S=1;$ the deficiency of the
reaction is 1.
\end{Exa}
Let us mention here that it is a boring task with many possibilities
of mistake to calculate the characteristic quantities of reactions
and this is one of the reasons why a program
package \textbf{\tt{ReactionKinetics.m}} is being developed in \textit{Mathematica},
see \cite{tothnagypapp}.
The second example may be prepared for the
present purposes as follows.

\begin{tabular}{ll}
\In{1}&\ttt{<\,<\,ReactionKinetics}${\ }\grave{ }$\\
\In{2}&\ttt{triangle =}
\ttt{ \{A}\(\rightleftharpoons\)\ttt{B}\(\rightleftharpoons\)\ttt{C}\(\rightleftharpoons\)\ttt{A\};}\\
\In{3}&\ttt{Column[ReactionsData[triangle]]}\\
&\ttt{species}\(\to\)\ttt{\{A,B,C\}}\\
&\(\mathfrak{M}\to\)\ttt{3}\\
&\ttt{externalspecies}\(\to\)\ttt{\{{\ }\}}\\
&\(\mathfrak{E}\to\)\ttt{0}\\
&\ttt{complexes}\(\to\)\ttt{\{A,B,C\}}\\
\Out{3}&\ttt{reactionsteps}\(\to\)
\ttt{\{A}\(\to\)\ttt{B,}
\ttt{B}\(\to\)\ttt{A,}
\ttt{B}\(\to\)\ttt{C,}
\ttt{C}\(\to\)\ttt{B,}
\ttt{C}\(\to\)\ttt{A,}
\ttt{A}\(\to\)\ttt{C\}}
\\
&\(\mathfrak{R}\to\)\ttt{6} \\
&\ttt{variables}\(\to\left\{\right.\)\tt{c}\mtt{$_A$},\tt{c}\mtt{$_B$},\tt{c}\mtt{$\left._C\right\}$}\\
&\(\alpha \to\)\mtt{\( \left(
\begin{array}{cccccc}
 1 & 0 & 0 & 0 & 0 & 1 \\
 0 & 1 & 1 & 0 & 0 & 0 \\
 0 & 0 & 0 & 1 & 1 & 0
\end{array}
\right)\)} \\
&\(\beta \to\)\mtt{\( \left(
\begin{array}{cccccc}
 0 & 1 & 0 & 0 & 1 & 0 \\
 1 & 0 & 0 & 1 & 0 & 0 \\
 0 & 0 & 1 & 0 & 0 & 1
\end{array}
\right)\)} \\
&\(\gamma \to\)\mtt{\( \left(
\begin{array}{rrrrrr}
-1 & 1  & 0  & 0  & 1  & -1 \\
 1 & -1 & -1 & 1  & 0  & 0 \\
 0 & 0  & 1  & -1 & -1 & 1
\end{array}
\right)\)} \\
\In{4}&\ttt{ShowFHJGraph[triangle,}
\ttt{\{k}$_{1}$, \ttt{k}$_{-1}$, \ttt{k}$_{2}$, \ttt{k}$_{-2}$, \ttt{k}$_{3}$, \ttt{k}$_{-3}$ \ttt{\},}
\\
&\ttt{VertexLabeling\ }\(\to\)\ttt{\ True},\ttt{\ DirectedEdges\ }\(\to\)\ttt{\ True]]}
\end{tabular}

\noindent Other uses of the package are described in the work
mentioned above.
\subsection{Models of ion channels}
In the models of ion channels the relevant species are
receptors and molecules modifying the operation of receptors
so as to change the sizes of the pores,
thereby decreasing or increasing the
quantity of ions flowing through the channels.
Altogether there are several hundreds of different
types of ion channels in living cells.

\begin{figure}[h!]
\centering
\includegraphics[width=0.8\textwidth]{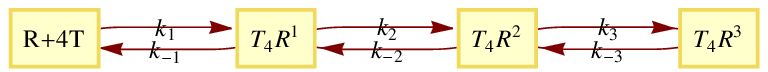}
\caption{The Érdi--Ropolyi model with four transmitters and three different states
of the transmitter-receptor complex}\label{fig:erdiropolyi}
\end{figure}

\begin{figure}[h!]
\centering
\includegraphics[width=0.6\textwidth]{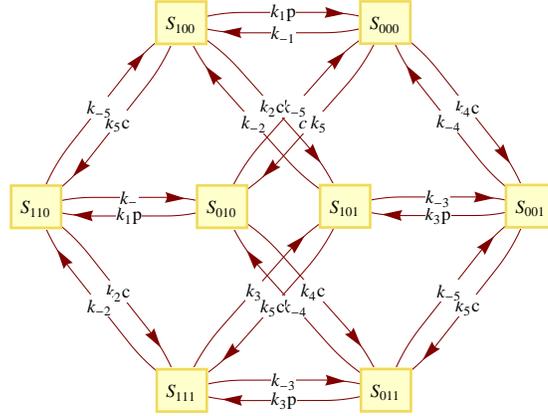}
\caption{A model by De Young and Keizer}\label{fig:deyoungkeizer}
\end{figure}

One possible model, see Fig. \ref{fig:erdiropolyi}, contains
receptors, transmitters, and receptor transmitter complexes each
with a different conformation having different ion-conductance, and
these conformations correspond to states in which the channels are
between the open and closed states \cite{erdiropolyi}.
Another approach, see Fig. \ref{fig:deyoungkeizer}, might involve
multiple types of modifying molecules and complexes, again
representing different states of the channels \cite{deyoungkeizer}.
These are the models we are especially interested in.

Ion channel models are usually required to fulfil
the principle of detailed balance both from theoretical
and practical points of view.
First, thermodynamics is said to require the principle to hold,
second, if this principle holds then the number of
reaction rate constants to be measured are reduced.
Let us turn to the formal definition of detailed balance
in the framework given in subsection \ref{subsec:frk}.
\subsection{Detailed balance: definition and the na\"{\i}ve approach}\label{db}
Within the model exactly defined above we can formulate
the property of being detailed balanced
\cite{hornjackson}.
Consider the reaction \eqref{reaction}
endowed with mass action kinetics.
\begin{Def}
If $\v{c_*}\in(\R^+)^M$ is such that
\begin{equation}
k_p\v{c_*}^{\alpha(.,p)}=k_{-p}\v{c_*}^{\beta(.,p)}
\quad(p=1,2,\dots,P),
\end{equation}
then reaction \eqref{reaction} is said to be \emph{detailed balanced
at the stationary point} $\v{c_*}.$ If the reaction is detailed
balanced at all its positive stationary points, then it is
\emph{detailed balanced}.
\end{Def}
We are especially interested in reactions
which are detailed balanced for some choices
of the reaction rate constants,
and also in the restrictions
upon the rate constants which ensure detailed balancing.
\begin{Exa}[Simple bimolecular reaction]
The deterministic model of the reaction $\revreak{A+B}{k_1}{k_{-1}}{C}$
according to subsection \ref{subsec:frk}
can be seen to be (in accord with the usual formulation)
\begin{center}
\begin{tabular}{lll}
\(a'=-k_1ab+k_{-1}c\)&\(b'=-k_1ab+k_{-1}c\)&\(c'=k_1ab-k_{-1}c\)\\
\(a(0)=a_0  \)&\(b(0)  =b_0 \)&\(c(0)=c_0\)\\
\end{tabular}
\end{center}
which simplifies to
\begin{eqnarray}
a'(t)&=&-k_1a(t)(a(t)-a_0+b_0)+k_{-1}(-a(t)+a_0+c_0)\nonumber\\
&=&-k_1a(t)^2+(k_1a_0-k_1b_0-k_{-1})a(t)+k_{-1}(a_0+c_0)\nonumber\\
&=&-k_{-1}\left(Ka(t)^2-(K(a_0-b_0)-1)a(t)-a_0-c_0\right).
\end{eqnarray}
If the reaction starts from nonnegative initial concentrations
\(a_0,b_0,c_0\) for which \(a_0+c_0>0,\)
the unique positive (relatively asymptotically stable)
equilibrium concentration
\begin{eqnarray*}
a_*&=&\frac{1}{2K}
(-1+K(a_0-b_0)+r)\\
b_*&=&\frac{1+K(a_0+b_0+2c_0)-r}{K(-1+K(a_0-b_0)+r)}\\
c_*&=&\frac{1}{2K}(1+K(a_0+b_0+2c_0)-r)\\
\mbox{\ where\ }K&:=&\frac{k_{1}}{k_{-1}},
r:=\sqrt{1+2K(a_0+b_0+2c_0)+K^2(a_0-b_0)^2}
\end{eqnarray*}
will be attained. The reaction is detailed balanced at this vector
of stationary concentrations for all values of the reaction rate
coefficients, i. e. $k_1a_*b_*=k_{-1}c_*$ always holds.
\end{Exa}
\begin{Exa}[Triangle reaction]
The induced kinetic differential equation of the reversible triangle
reaction being
\begin{eqnarray*}
a'&=&-k_1a+k_{-1}b-k_{-3}a+k_{3}c\\
b'&=&k_1a-k_{-1}b-k_2b+k_{-2}c\\
c'&=&k_2b-k_{-2}c+k_{-3}a-k_{3}c
\end{eqnarray*}
together with the mass conservation relation
\begin{eqnarray*}
a(t)+b(t)+c(t)=a_0+b_0+b_0=:m
\end{eqnarray*}
imply that the unique, relatively asymptotically stable  vector of
positive stationary concentrations---if
at least one of the initial concentrations \(a_0,b_0,c_0\) is positive---
are as follows.
\begin{eqnarray}
a_*&=&(k_{-2} k_{-1}+(k_{-1}+k_2) k_3)\frac{ m}{d}\label{trstac1}\\
b_*&=&(k_{-3} k_{-2}+(k_{-2}+k_3)k_1)\frac{ m}{d}\label{trstac2}\\
c_*&=&(k_{-3} k_{-1}+(k_{-3}+k_1)k_2)\frac{ m}{d}\label{trstac3}\\
\mbox{with }d&:=&k_{-2} (k_{-1}+k_1)+k_1k_2+k_{-3} (k_{-2}+k_{-1}+k_2)+k_{-1} k_3+k_1 k_3+k_2 k_3.\nonumber
\end{eqnarray}
The reaction is detailed balanced at this
vector of stationary concentrations---i. e.
\[
k_1a_*=k_{-1}b_*\quad
k_2b_*=k_{-2}c_*\quad
k_3c_*=k_{-3}a_*
\]
---if and only if
\begin{equation}\label{tricircle}
k_1k_2k_3=k_{-1}k_{-2}k_{-3}
\end{equation}
holds.
\end{Exa}
\begin{Exa}[Wegscheider]
The induced kinetic differential equation of the Wegscheider reaction being
\begin{eqnarray*}
a'&=&-k_1a+k_{-1}b-k_2a^2+k_{-2}ab\\
b'&=&k_1a-k_{-1}b+k_2a^2-k_{-2}ab
\end{eqnarray*}
---which simplifies to
\begin{eqnarray*}
a'&=&-k_1a+k_{-1}(a_0+b_0-a)-k_2a^2
+k_{-2}a(a_0+b_0-a)\nonumber\\
&=&-(k_2+k_{-2})a^2-(k_1+k_{-1}-k_{-2}(a_0+b_0))a+k_{-1}(a_0+b_0).
\end{eqnarray*}
---together with the mass conservation relation
\begin{eqnarray*}
a(t)+b(t)=a_0+b_0=:m
\end{eqnarray*}
imply that---unless all the initial concentrations are zero---the
unique positive (relatively asymptotically stable)
stationary concentration vector is as follows.
\begin{eqnarray}\label{wegstac1}
a_*&=&\frac{k_{-1}+k_1-k_{-2} m-r}{-2(k_{-2}+k_2)}\\
b_*&=&\frac{k_{-1}+k_1+k_{-2} m+2k_2m-r}{2(k_{-2}+k_2)}\label{wegstac2}\\
\mbox{with }r&:=&\sqrt{(k_{-1}+k_1-k_{-2}m)^2+4k_{-1}m(k_{-2}+k_2)}.\label{appstac13}
\end{eqnarray}
The reaction is detailed balanced at this
vector of stationary concentrations---i. e.
\[
k_1a_*=k_{-1}b_*,k_2a_*b_*=k_{-2}b_*^2
\] ---if and only if
\begin{equation}\label{wegspan}
\frac{k_{1}}{k_{-1}}=\frac{k_{2}}{k_{-2}}
\end{equation}
holds.
\end{Exa}
\subsection{The necessary and sufficient condition of detailed balancing}
The necessary and sufficient conditions are formulated in the
following way in \cite{feinberg}. Consider the reaction
\eqref{reaction} endowed with mass action kinetics.

First suppose that we have chosen an arbitrary
spanning forest for the FHJ-graph of the network.
It is possible to find a set of $P-(N-L)$ independent
circuits induced by the choice of the spanning forest.
For each of these circuits we write an equation
which asserts that the product of the rate constants
in the clockwise direction and the counterclockwise
direction is equal.
Thus we have $P-(N-L)$ equations:
the \emph{circuit conditions}.

Next, these equations are supplemented with the $\delta$
\emph{spanning forest conditions} as follows.
Suppose that the edges of the spanning forest has
been given an orientation.
Then there are $\delta$ independent
nontrivial solutions to the vector equation
$\sum_{(i,j)}a_{ij}\v{v}_{ij}=\v{0}$
where the sum is taken for
all reaction steps in the oriented spanning forest and $\v{v}_{ij}$
is the corresponding reaction step vector.
With these $a_{ij}$ coefficients the
spanning forest conditions are
\begin{equation}
\prod k_{ij}^{a_{ij}}=\prod k_{ji}^{a_{ij}},
\end{equation}
where $k_{ij}$ are the corresponding rate coefficients.

With all these the widely-accepted \emph{necessary}
conditions (the  circuit conditions)
are complemented with the
spanning forest conditions to form a set of
\emph{necessary and sufficient} conditions for
detailed balancing in
mass action systems of arbitrary complexity.
\begin{Thm}[Feinberg]
The reaction \eqref{reaction} is detailed balanced
for all those choices of the reaction rate constants
which satisfy the $P-(N-L)$ circuit conditions and the $\delta$
spanning forest conditions.
\end{Thm}

\begin{Rem}
The circuit conditions are called \emph{spanning tree method} in
\cite{colquhoun1}.
\end{Rem}

\begin{Rem}
There are three interesting special cases.
\begin{enumerate}
\item
For a reversible mass action system which has a deficiency of zero,
the circuit conditions \emph{alone} become necessary and sufficient
for detailed balancing. The reason why the circuit conditions were
generally accepted as sufficient as well, is that a large majority
of models are of zero deficiency. This case is exemplified by the
triangle reaction.
\item
For networks with no nontrivial circuits, that is, in which there
are just $N-L$ reaction pairs and so $P-(N-L)=0$, the circuit
conditions are vacuous. Therefore, the spanning forest conditions
\emph{alone} are necessary and sufficient for detailed balancing.
The example by Wegscheider belongs to this category.
\item
Finally, if a reversible network is circuitless and has
a deficiency of zero,
both the circuit conditions and the spanning forest
conditions are vacuous.
The system is detailed balanced
(or fulfils the principle of microscopic reversibility),
regardless of the values of the rate constants.
Such is a compartmental system with no circles in
the FHJ-graph, the simple bimolecular reaction or
the Érdi--Ropolyi model.
\end{enumerate}
\end{Rem}

\section{The main result}
\subsection{Our strategy}

Let us denote by $M, P, \delta, N, L, S, K$ and $M', P', \delta',
N', L', S', K'$ the number of species, the number of (half) reaction
steps, the deficiency, the number of complexes, the number of
linkage classes, the dimension of the stoichiometric space (i.e.,
the number of independent reaction steps) and the number of
independent cycles respectively in the original and in the
transformed system.

All the investigated original (not mass-conserving) ion channel
models are formally compartmental systems which means that each
complex consists of a single species and all species are different.
Therefore all these models are of deficiency zero. Thus, in order to
check detailed balancing it is enough to test the circuit
conditions, and this is what the authors in
\cite{colquhoun2,colquhoun1} do.

What we propose is to transform these models into a mass-conserving
model in such a way as to reflect the same physical reality. The
transformed models have the following properties.
\begin{enumerate}
\item
There is no cycle in the transformed system.
\item $S=S'$
\item
$N'-L'-S'=\delta'=K$
\item
The circuit conditions in the original system are
equivalent to the spanning forest conditions in the transformed
system.
\end{enumerate}

This transformation is constructed in the Appendix
for a large class of systems---those
with rectangular grids as FHJ-graphs---containing
all the special cases we have met up to now.

\subsection{Lemma}
Consider a directed graph whose edges and vertices are the edges and
vertices of a planar rectangular grid. Suppose that the graph has
$n$ vertices and that to each vertex $j$ we assign a $\v{y}_j$
vector in $\mathbf{R}^{n+2}$ such that these vertex vectors are
linearly independent. Let $\v{c}_1$ and $\v{c}_2$ be vectors in
$\mathbf{R}^{n+2}$ such that they are linearly independent of each
other and of each $\v{y}_j$. Let us denote by $e_{ij}$ the directed
edge of the graph from vertex $i$ to vertex $j$ and to each
$e_{ij}$ edge let us assign the $\v{v}_{ij}=\v{y}_j-\v{y}_i$ vector.
Let us define the $\v{u}_{ij}$ vectors in the following way.
\begin{figure}[h!]
  \centering
  \subfloat[][]{\label{fig:lemma1}\includegraphics[width=0.465\textwidth]{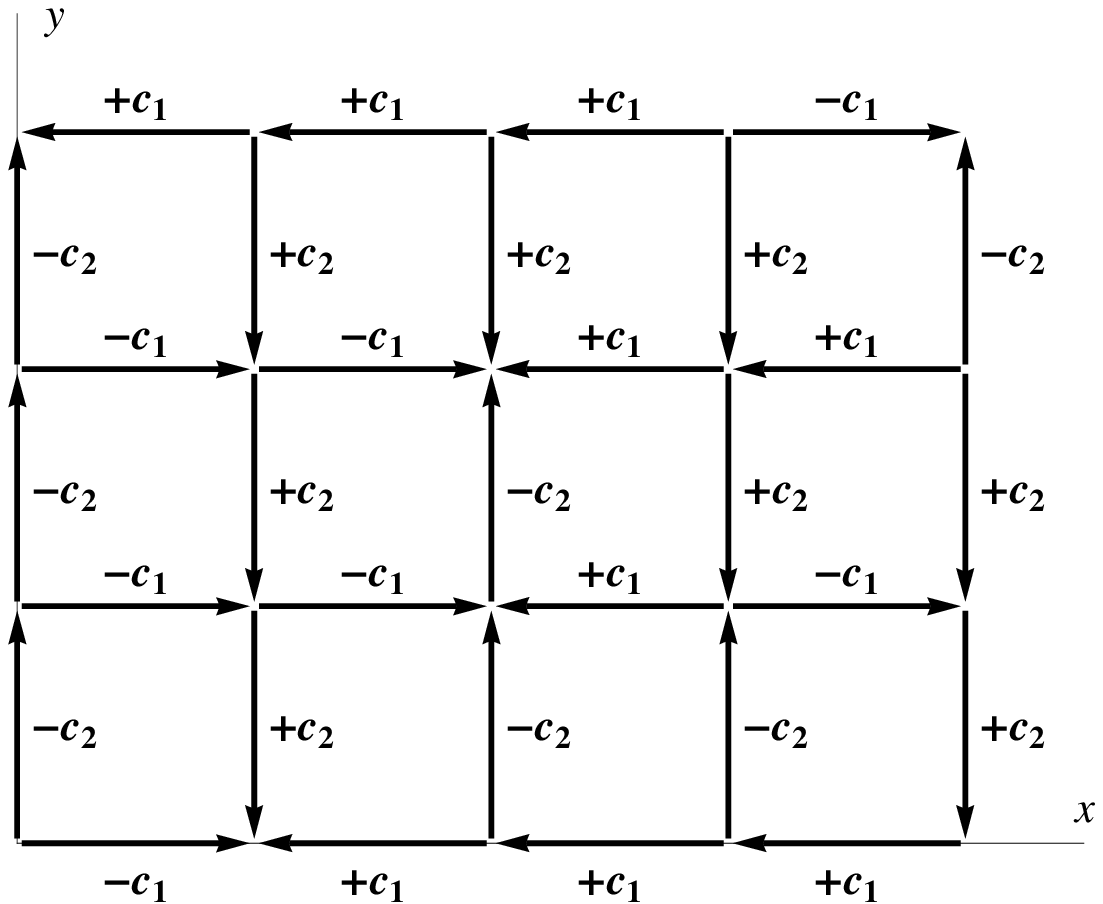}}\qquad
  \subfloat[][]{\label{fig:lemma2}\includegraphics[width=0.465\textwidth]{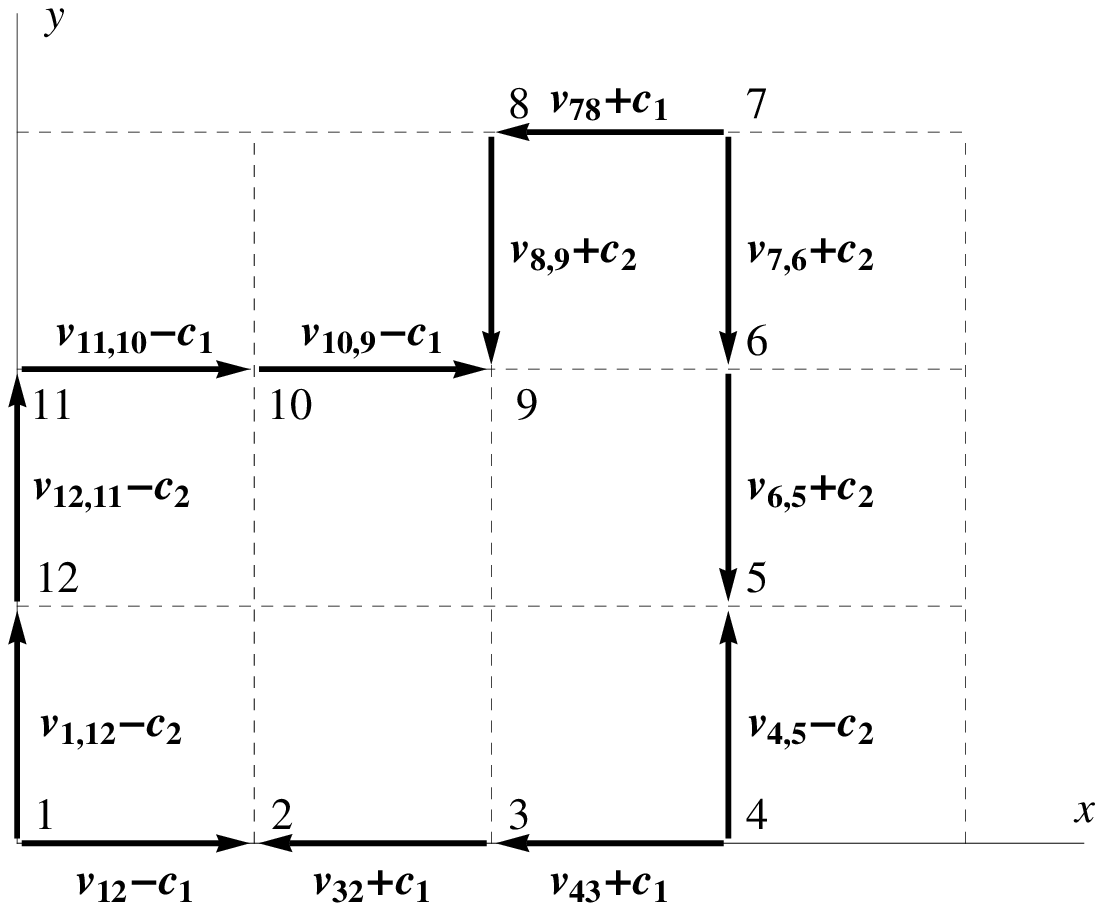}}
  \caption{}
  \label{fig:lemma}
\end{figure}
If $e_{ij}$ is directed in the positive or negative direction in
relation to the $x$ axis then $\v{u}_{ij}:=\v{v}_{ij}-\v{c}_1$ or
$\v{u}_{ij}:=\v{v}_{ij}+\v{c}_1$, respectively. Similarly, if $e_{ij}$
is directed in the positive or negative direction in relation to the
$y$ axis then $\v{u}_{ij}:=\v{v}_{ij}-\v{c}_2$ or
$\v{u}_{ij}:=\v{v}_{ij}+\v{c}_2$, respectively. Let us denote by
span$\{\v{v}_{ij}\}$ the subspace generated by the $\v{v}_{ij}$
vectors.

\begin{Lem} Under these conditions the following statements hold.
\begin{enumerate}
\item Along each directed circle in the graph,
$\sum a_{ij}\v{v}_{ij}=\sum a_{ij}\v{u}_{ij}=\v{0}$ where $a_{ij}:=1$ if
the edges of the graph and the circle are directed in the same way
and $a_{ij}:=-1$ otherwise.
\item The dimension of span$\{\v{v}_{ij}\}$ and span$\{\v{u}_{ij}\}$ is $n-1$.
\end{enumerate}
\end{Lem}

\textbf{Proof} 1. Since the $\v{v}_{ij}$ vectors are the
differences of the corresponding vertex vectors, it is obvious that
along a directed circle, the sum of the $\v{v}_{ij}$ vectors
is~$\v{0}$. It is enough to show that the $\v{c}_1$ and $\v{c}_2$
vectors disappear in the sum of the $\v{u}_{ij}$ vectors. In order
to see this, first assume that along a directed circle we change the
direction of the $e_{ij}$ edges so that each is directed
clockwise. In this case it is obvious that the sum of the $\v{c}_1$
and $\v{c}_2$ vectors is zero since the number of the "$+\v{c}_1$" and "$+\v{c}_2$"
vectors is equal to the number of the "$-\v{c}_1$" and "$-\v{c}_2$" vectors, respectively.
Then, changing the original directions back, the sign of the $\v{c}_1$ and $\v{c}_2$ vectors changes twice and
thus they will not appear in the sum.
\\
2. Let us choose a spanning tree in the graph consisting of $n-1$ of
the $e_{ij}$ edges. Then the corresponding $\v{v}_{ij}$ vectors are
linearly independent and since the $\v{c}_1$ and $\v{c}_2$ vectors are
independent of them, the corresponding $\v{u}_{ij}$ vectors are also
linearly independent.

\begin{Rem} It is trivial that the statements of the lemma
remain true if either $\v{c}_1$ or $\v{c}_2$ is the zero vector, or, if
the graph contains edges that are not part of a circle.
\end{Rem}
\begin{Rem}
The statements of the lemma are also true for graphs consisting of $k$-dimensional grids ($k\geq3$), see Fig. \ref{fig:system3} and Fig. \ref{fig:system4a}, \ref{fig:system4b} as an illustration for the three-dimensional case.
\end{Rem}

\subsection{Examples}
In the next three examples, the left side of the figure shows the original system and the right side of the figure
shows the transformed system with an oriented spanning forest. Both systems are
reversible, the arrows show a direction needed to write down the spanning forest conditions.
The choice of the numbering of the species as well as the direction of the reaction vectors
is arbitrary but in both systems they are chosen correspondingly.

\begin{Exa}\label{pelda1}
The system in Fig. \ref{fig:system1} can be found in \cite{colquhoun2}.
The meaning of the species is as follows: The core of the system is
obviously a rectangle, the additional parts do not mean an extra
problem as the reader can easily verify it.
\begin{figure}[h!]
  \centering
  \subfloat[][$M=10, N=10, L=1, S=9,$\\ $\delta=0,
  K=2$]{\label{fig:system1a}\includegraphics[width=0.39\textwidth]{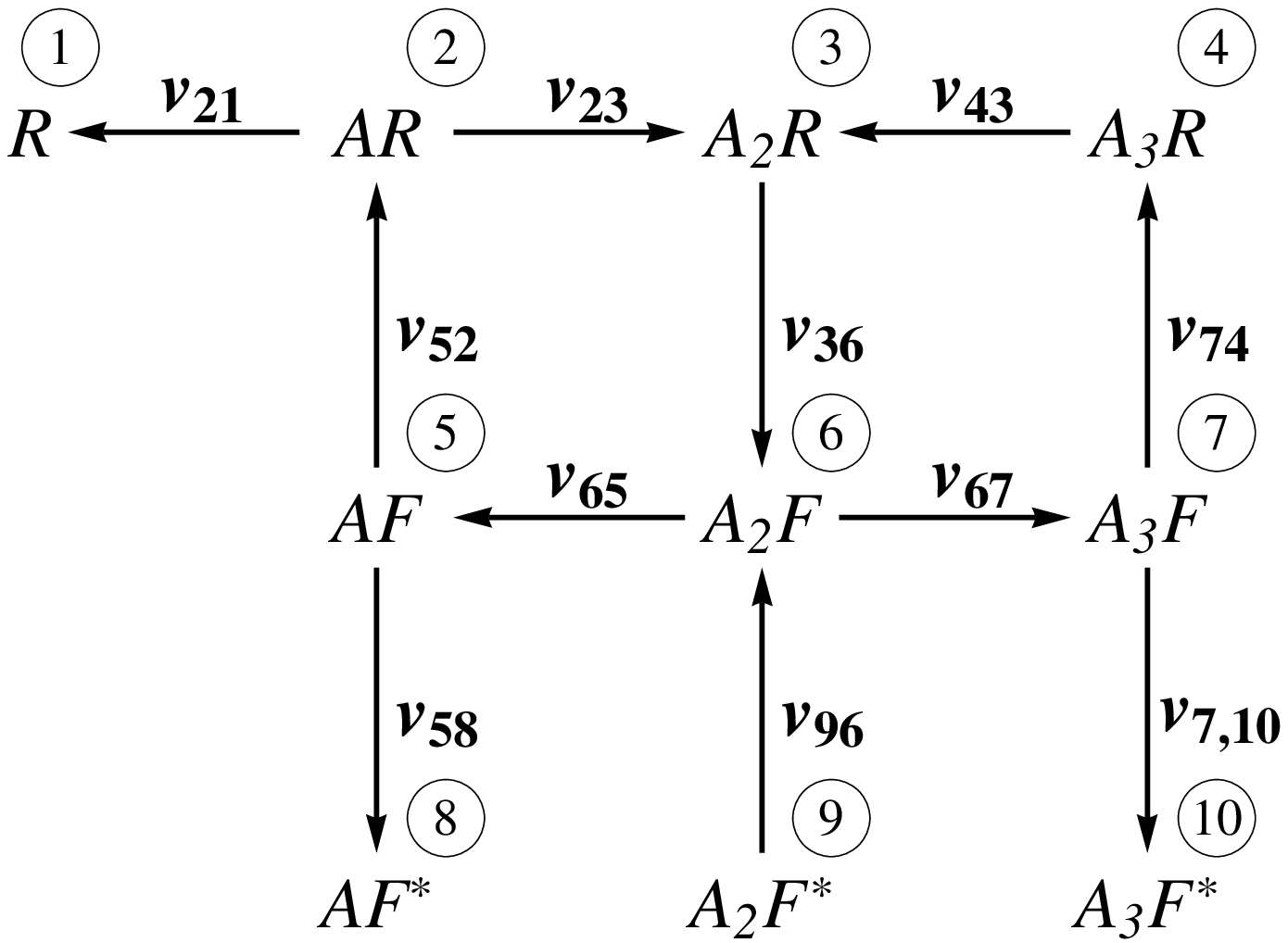}}\quad
  \subfloat[][$M'=11, N'=14, L'=3, S'=9,$\\ $\delta'=2, K'=0$]{\label{fig:system1b}\includegraphics[width=0.57\textwidth]{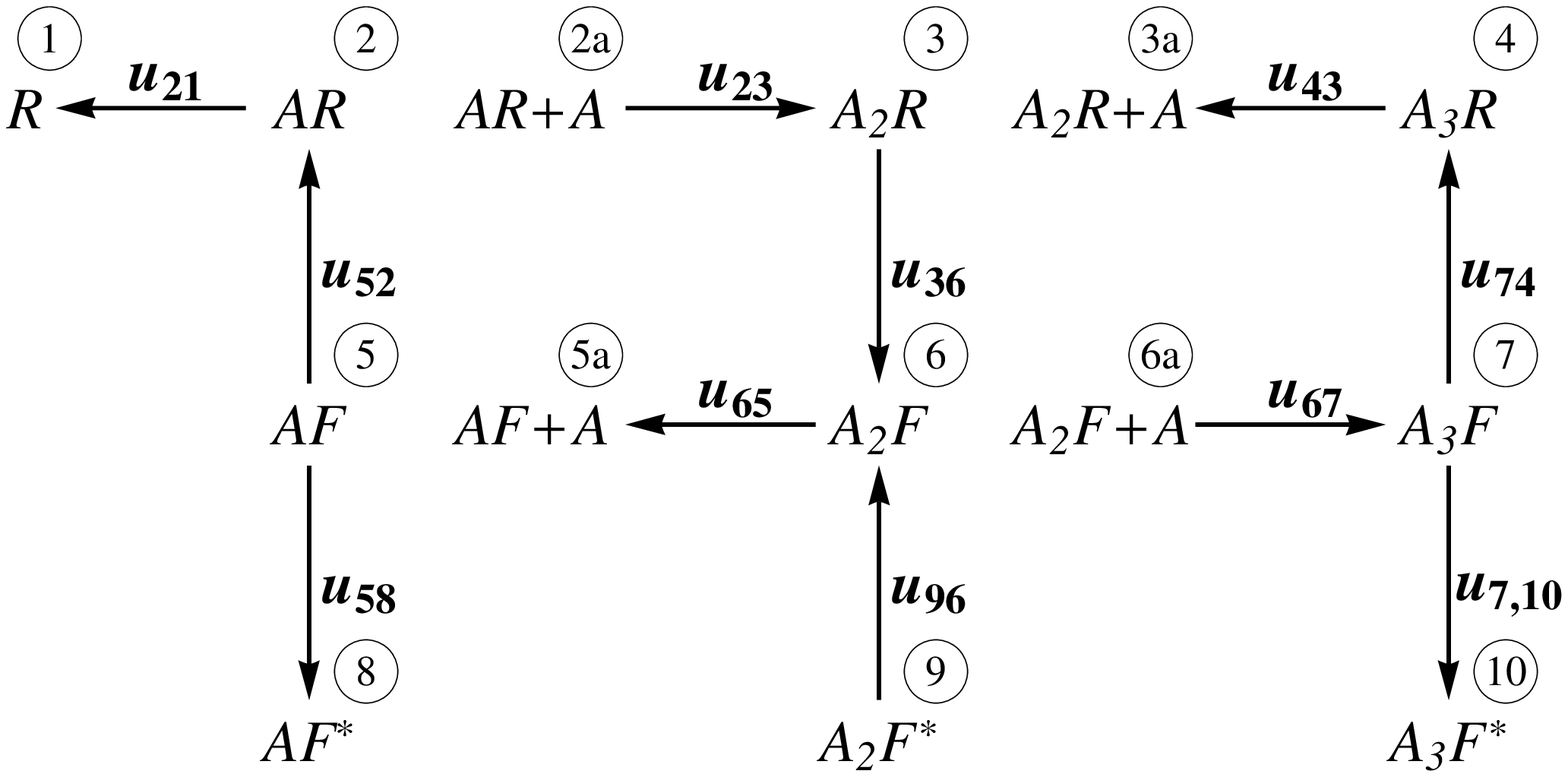}}
  \caption{}\label{fig:system1}
\end{figure}
The original system consists of $M=10$ species, $N=10$ complexes,
$L=1$ linkage class and it contains two circles while the
transformed system contains one more species, $A$, there are $N'=14$
complexes, $L'=3$ linkage classes and it is circuitless. In order to
compare these systems easily, in both cases let us number the
species in the same way and let $A$ be the last one, that is,
$$
X(1):=R, X(2):=AR, X(3):=A_2R,\dots,~X(10):=A_3F^*, X(11):=A.
$$
Let us assign a vector $\v{y}_i\in\R^{11}$ to the $i$th species so that
$y_{i,j}=1$ if $i=j$ and $y_{i,j}=0$ if $i\neq j$ where
$i,j=1,\dots,11$ and let $\v{a}:=\v{y}_{11}$.

The complex vectors in Fig. \ref{fig:system1a} are $\v{y}_1,
\v{y}_2,\dots, \v{y}_{10}$ and the corresponding reaction vectors
are $\v{v}_{21}=\v{y}_1-\v{y}_2, \v{v}_{23}=\v{y}_3-\v{y}_2,\dots,
\v{v}_{7,10}=\v{y}_{10}-\v{y}_7$. The dimension of
span$\{\v{v}_{21}, \v{v}_{23}, \dots,\v{v}_{7,10}\}$ is $S=9$. Thus,
the deficiency of this system is $\delta=N-L-S=0$. It means that the
circuit conditions are necessary and sufficient for detailed
balancing. The circuit conditions along circles 2365 and 4367 are
\begin{eqnarray*}
k_{23}k_{36}k_{65}k_{52}=k_{32}k_{25}k_{56}k_{63}\\
k_{43}k_{36}k_{67}k_{74}=k_{34}k_{47}k_{76}k_{63}
\end{eqnarray*}

The complexes in Fig. \ref{fig:system1b} are numbered as $1, 2, 2a,
\dots, 10$ and the complex vectors are $\v{y}'_1=\v{y}_1$,
$\v{y}'_2=\v{y}_2$, $\v{y}'_{2a}=\v{y}_2+\v{a}$, \dots,
$\v{y}'_{10}=\v{y}_{10}$. The reaction vectors are
$\v{u}_{21}=\v{v}_{21}$, $\v{u}_{23}=\v{v}_{23}-\v{a}$,
$\v{u}_{43}=\v{v}_{43}+\v{a}$, $\v{u}_{52}=\v{v}_{52}$,
$\v{u}_{36}=\v{v}_{36}$, $\v{u}_{74}=\v{v}_{74}$,
$\v{u}_{65}=\v{v}_{65}+\v{a}$, $\v{u}_{67}=\v{v}_{67}-\v{a}$,
$\v{u}_{58}=\v{v}_{58}$, $\v{u}_{96}=\v{v}_{96}$,
$\v{u}_{7,10}=\v{v}_{7,10}$. The lemma can be applied to this system
with $\v{c}_1=\v{a}$ and $\v{c}_2=\v{0}$. The dimension of span$\{\v{u}_{21}, \v{u}_{23},
\dots,\v{u}_{7,10}\}$ is also $S'=9$. Thus, the deficiency is
$\delta'=14-3-9=2$. Since this system is circuitless, there are two
equations according to the spanning forest conditions that ensure
detailed balancing. Along the circles '$2365$' and '$3476$' in
both systems,
$\v{v}_{23}+\v{v}_{36}+\v{v}_{65}+\v{v}_{52}=\v{u}_{23}+\v{u}_{36}+\v{u}_{65}+\v{u}_{52}=\v{0}$
and
$\v{v}_{43}+\v{v}_{36}+\v{v}_{67}+\v{v}_{74}=\v{u}_{43}+\v{u}_{36}+\v{u}_{67}+\v{u}_{74}=\v{0}$.
Since each coefficient of the $\v{u}_{ij}$ vectors in the above
linear combinations is 1,
\begin{eqnarray*}
k'_{23}k'_{36}k'_{65}k'_{52}=k'_{32}k'_{63}k'_{56}k'_{25}\\
k'_{43}k'_{36}k'_{67}k'_{74}=k'_{34}k'_{63}k'_{76}k'_{47}
\end{eqnarray*}
which are equivalent to the circuit conditions.
\end{Exa}

\begin{Rem}
Let us observe that the equivalence of the circuit conditions in the original system and the spanning forest conditions in the transformed system follows from the first statement of the lemma, that is, along each circle the $\v{v}_{ij}$ vectors and the correspondingly chosen $\v{u}_{ij}$ vectors satisfy the same linear equalities. If, say, instead of circle '$4367$' we choose circle '$234765$' then
$\v{v}_{23}-\v{v}_{43}-\v{v}_{74}-\v{v}_{67}+\v{v}_{65}+\v{v}_{52}=\v{u}_{23}-\v{u}_{43}-\v{u}_{74}-\v{u}_{67}+\v{u}_{65}+\v{u}_{52}=\v{0}$.
The corresponding circuit condition in Fig. \ref{fig:system1a} is
\[
k_{23}k_{34}k_{47}k_{76}k_{65}k_{52}=k_{32}k_{25}k_{56}k_{67}k_{74}k_{43}
\]
and the equivalent equation from the spanning forest condition in
Fig. \ref{fig:system1b} is
$k'_{23}(k'_{43})^{-1}(k'_{74})^{-1}(k'_{67})^{-1}k'_{65}k'_{52}=k'_{32}(k'_{34})^{-1}(k'_{47})^{-1}(k'_{76})^{-1}k'_{56}k'_{25}$.
\end{Rem}

\begin{Exa}\label{pelda2}
The system in Fig. \ref{fig:system2} can be found in \cite{nahum-levy}. Fig.
\ref{fig:system2a} shows the original system where there are $M=11$
species and Fig. \ref{fig:system2b} shows the transformed system
where there are two more species, $A$ and $G$. Again, let us number
the species in the same way as in Fig. \ref{fig:system1} and let $A$
and $G$ be the last two, that is,
$$
X(1):=R, X(2):=RA,\dots, X(11):=G_2R'A_2, X(12):=A, X(13):=G.
$$

\begin{figure}[h!]
  \centering
  \subfloat[][$M=11, N=11, L=1,$\\ $S=10, \delta=0,
  K=4$]{\label{fig:system2a}\includegraphics[width=0.39\textwidth]{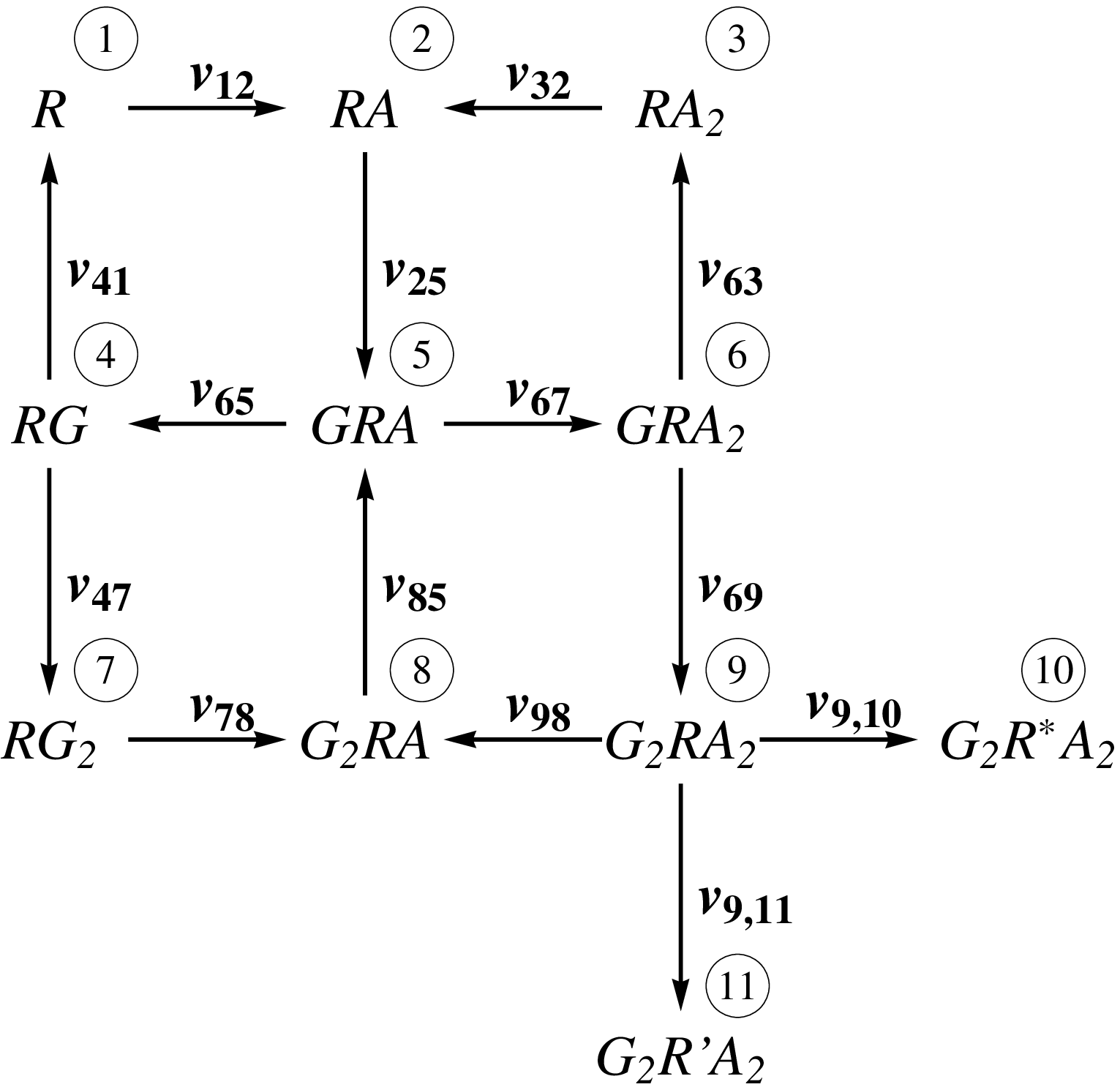}}\quad
  \subfloat[][$M'=13, N'=22, L'=8,$\\ $S'=10,\delta'=4, K'=0$]{\label{fig:system2b}\includegraphics[width=0.57\textwidth]{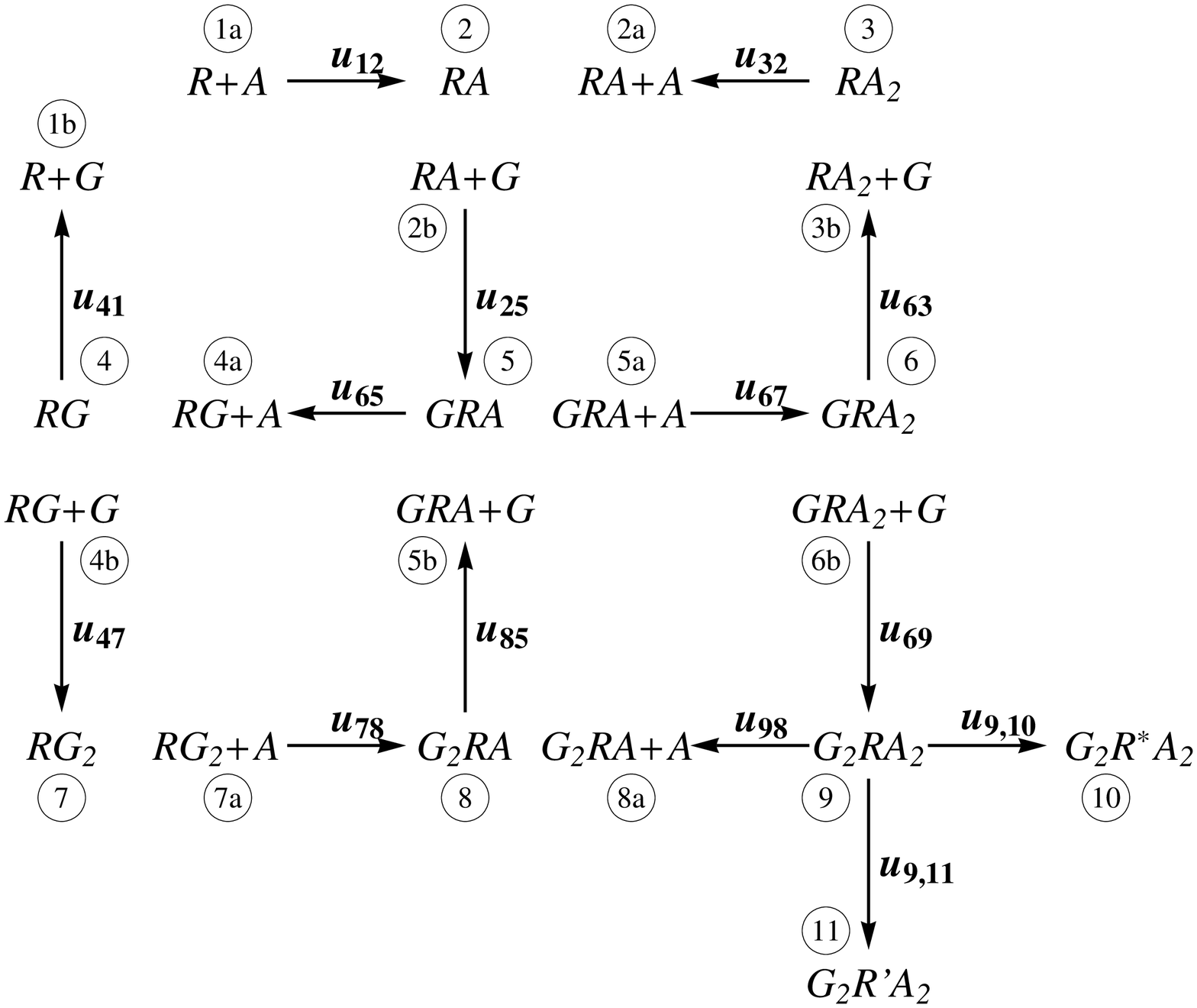}}
  \caption{}\label{fig:system2}
\end{figure}

Let us assign a vector $\v{y}_i\in\R^{13}$ to the $i$th species so
that $\v{y}_{i,j}=1$ if $i=j$ and $\v{y}_{i,j}=0$ if $i\neq j$ where
$i,j=1,\dots,13$. With $\v{a}:=\v{y}_{12}$ and $\v{g}:=\v{y}_{13}$,
the corresponding reaction vectors of the transformed system are
$\v{u}_{12}=\v{v}_{12}-\v{a}$, $\v{u}_{32}=\v{v}_{32}+\v{a}$,
$\v{u}_{41}=\v{v}_{41}+\v{g}$, $\v{u}_{25}=\v{v}_{25}-\v{g}$, \dots,
$\v{u}_{98}=\v{v}_{98}+\v{a}$, $\v{u}_{9,10}=\v{v}_{9,10}$,
$\v{u}_{9,11}=\v{v}_{9,11}$. Thus, the lemma can be applied so that
$\v{c}_1=\v{a}$, $\v{c}_2=\v{g}$ and the $x$ and $y$ axes are directed in
the '147' and '123' direction, respectively.
\end{Exa}

\begin{Exa}\label{pelda3}

The system in Fig. \ref{fig:system3} can be found in \cite{colquhoun3}.
\begin{figure}[h!]
  \centering
  \subfloat[][$M=18, N=18, L=1, S=17,$\\ $\delta=0,
  K=13$]{\label{fig:system3a}\includegraphics[width=0.43\textwidth]{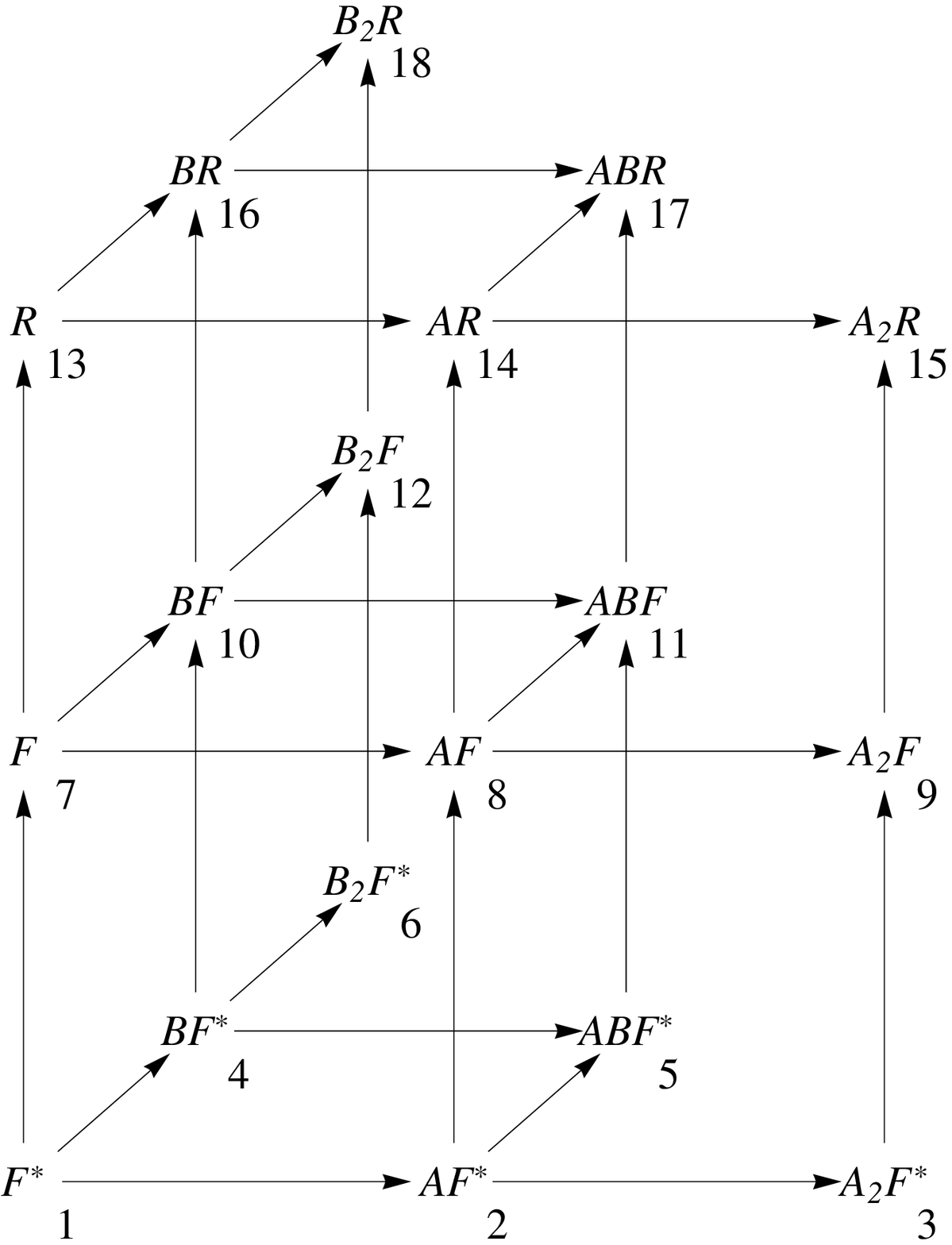}}\qquad
  \subfloat[][$M'=20, N'=36, L'=6, S'=17,$\\ $\delta'=13, K'=0$]{\label{fig:system3b}\includegraphics[width=0.5\textwidth]{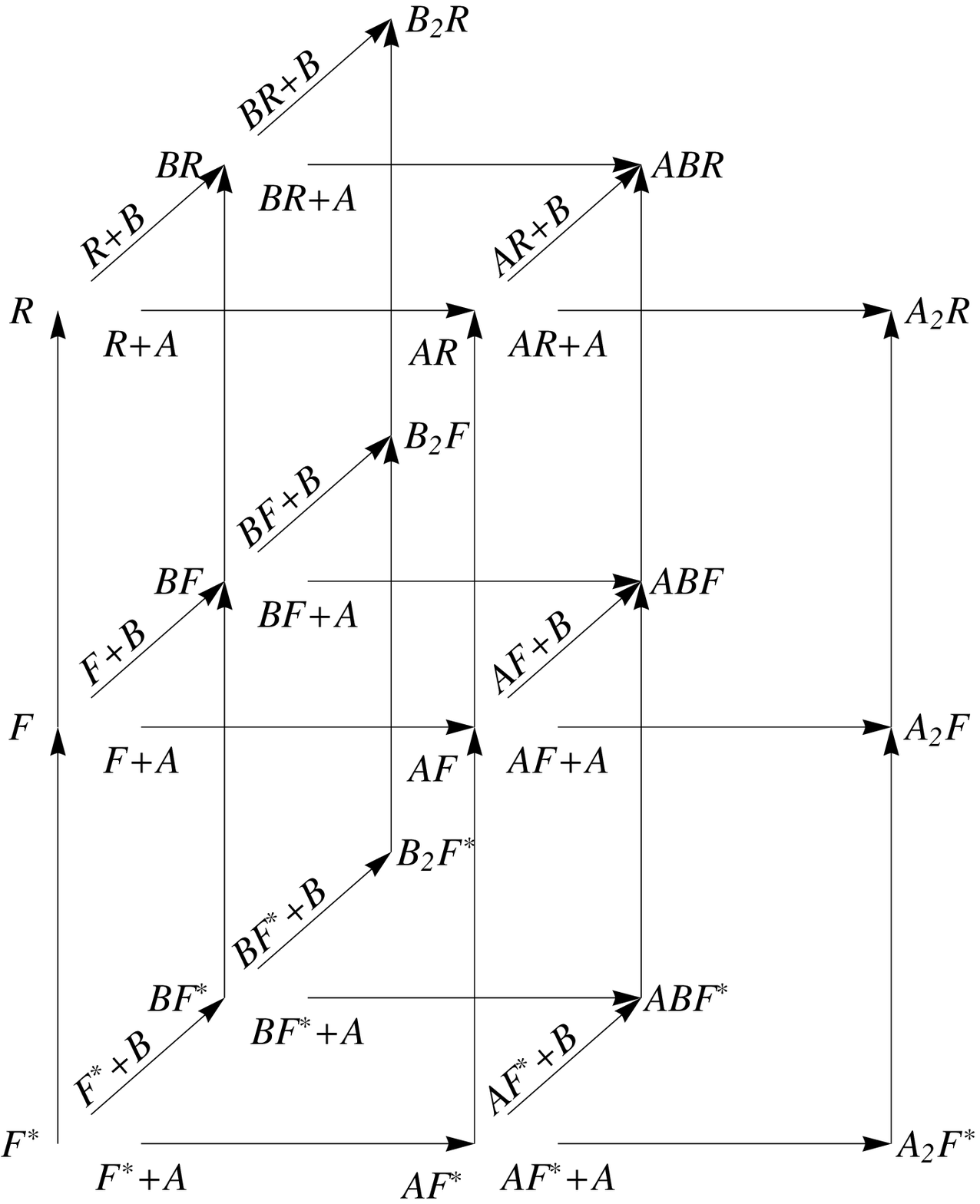}}
  \caption{}\label{fig:system3}
\end{figure}
Again, let us number the species as shown in Fig. \ref{fig:system3a}, that is
$$
X(1):=F^*, X(2):=AF^*,\dots, X(18):=B_2R, X(19):=A, X(20):=B.
$$
and let us assign a vector $\v{y}_i\in\R^{20}$ to the $i$th species so
that $\v{y}_{i,j}=1$ if $i=j$ and $\v{y}_{i,j}=0$ if $i\neq j$ where
$i,j=1,\dots,20$. Similarly as in the previous two cases, let $\v{v}_{ij}$ and $\v{u}_{ij}$ respectively denote the reaction vectors
in the original and in the transformed system (these are the differences of the corresponding complex vectors) and let $\v{a}:=\v{y}_{19}$ and $\v{b}:=\v{y}_{20}$. Then, $\v{u}_{ij}=\v{v}_{ij}-\v{a}$, $\v{u}_{ij}=\v{v}_{ij}-\v{b}$ or $\v{u}_{ij}=\v{v}_{ij}$ if $\v{u}_{ij}$ and $\v{v}_{ij}$ correspond to an edge in the graph parallel to the '$12$', '$14$' or '$17$' directions, respectively. The lemma can be used here with
$\v{c}_1=\v{a}$, $\v{c}_2=\v{b}$ and $\v{c}_3=\v{0}$.

\end{Exa}

\begin{Exa}\label{pelda4} Consider the system in Fig. \ref{fig:system4a} where
there is one receptor with three binding sites,
and the different states of the sites are denoted by $S_{ijk}$.
The next three figures show three possible transformation of this system in the following cases.
Fig. \ref{fig:system4b} shows the transformed versions of the system in Fig. \ref{fig:system4a} in the case when there are three different atoms, $A$, $B$, $C$, binding to the three sites. Fig. \ref{fig:system4c} shows the transformed version of the system in Fig. \ref{fig:system4a} in the case when there are two different atoms, $A$, $B$, binding to the three sites. This is the De Young and Keizer model, and again, the transformed system does not contain a circle. In the interesting theoretical case when there is only one atom, $A$, binding to each of the three sites, the transformed system contains a circle, this can be seen in Fig. \ref{fig:system4c}. It can be verified easily that the five circuit conditions in the original system are equivalent to the five spanning forest conditions in the systems in Fig. \ref{fig:system4b} and \ref{fig:system4c} and are also equivalent to the four spanning forest conditions and one circuit condition in the system in Fig. \ref{fig:system4d}.

\begin{figure}[h!]
  \centering
  \subfloat[][$M=8, N=8, L=1,$\\$S=7, \delta=0, K=5$]{\label{fig:system4a}\includegraphics[width=0.35\textwidth]{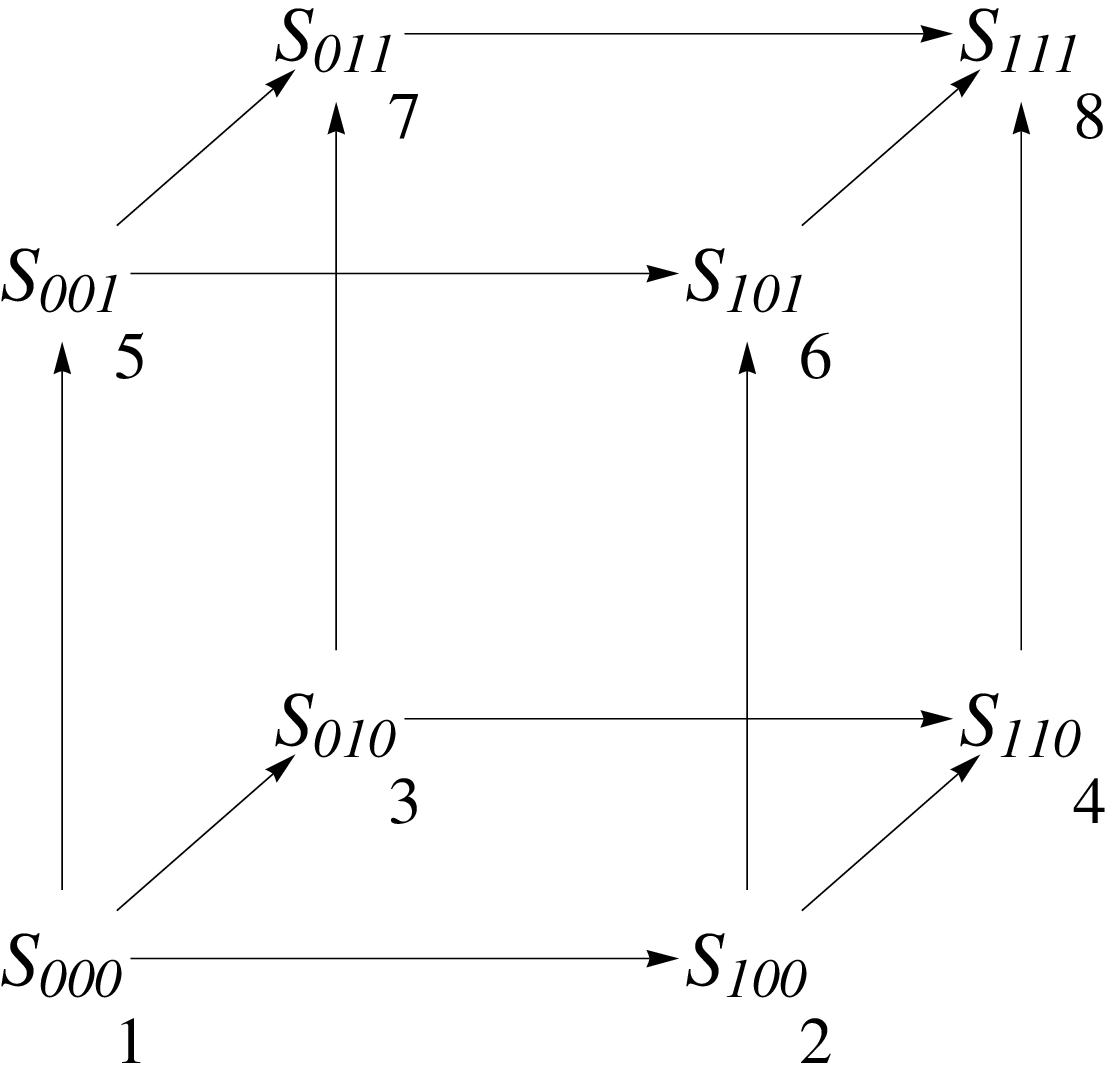}}\qquad\qquad
  \subfloat[][$M'=11, N'=19, L'=7,$\\$S'=7, \delta'=5, K'=0$]{\label{fig:system4b}\includegraphics[width=0.4\textwidth]{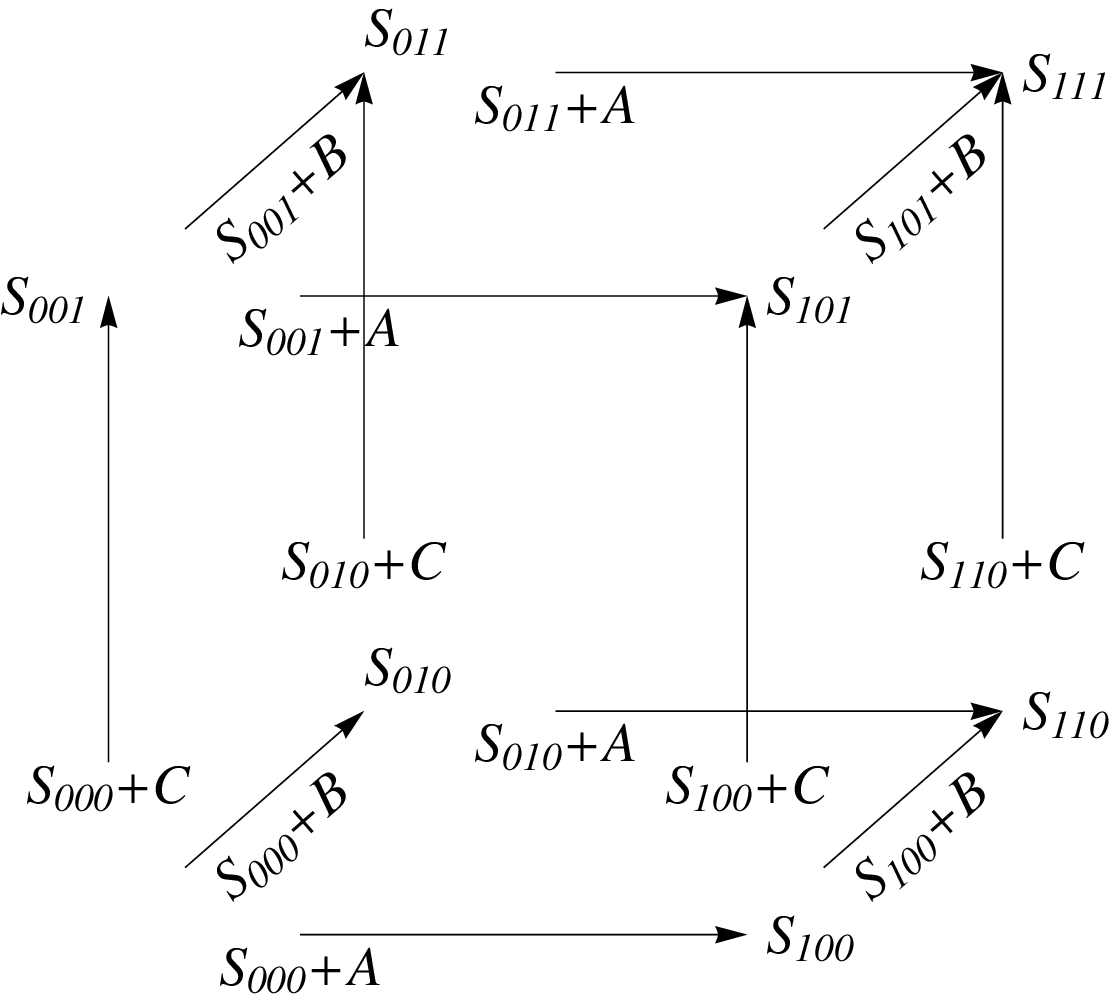}}\qquad
  \subfloat[][$M'=10, N'=17, L'=5,$\\$S'=7, \delta'=5, K'=0$]{\label{fig:system4c}\includegraphics[width=0.35\textwidth]{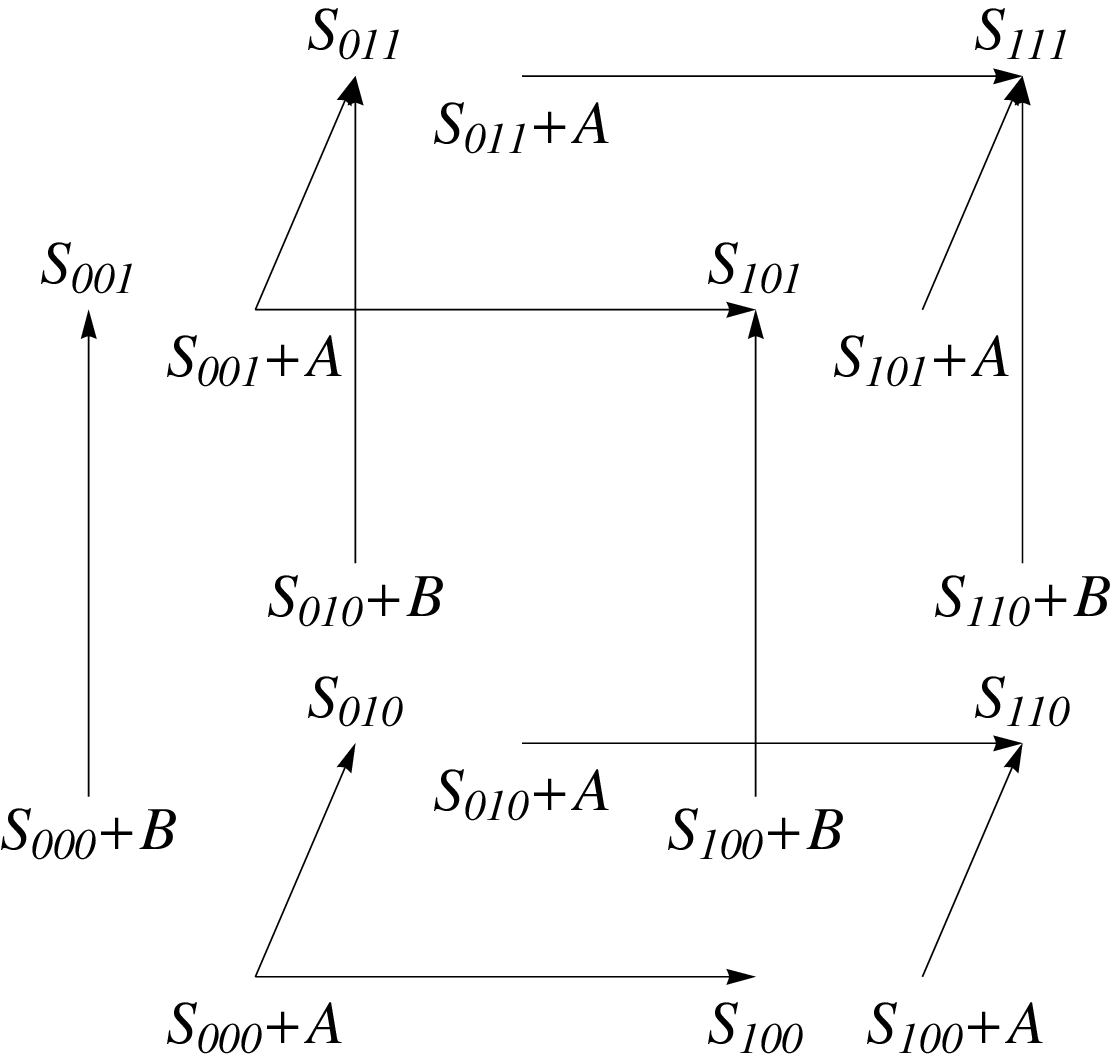}}\qquad\qquad
  \subfloat[][$M'=9, N'=14, L'=3,$\\$S'=7, \delta'=4, K'=1$]{\label{fig:system4d}\includegraphics[width=0.4\textwidth]{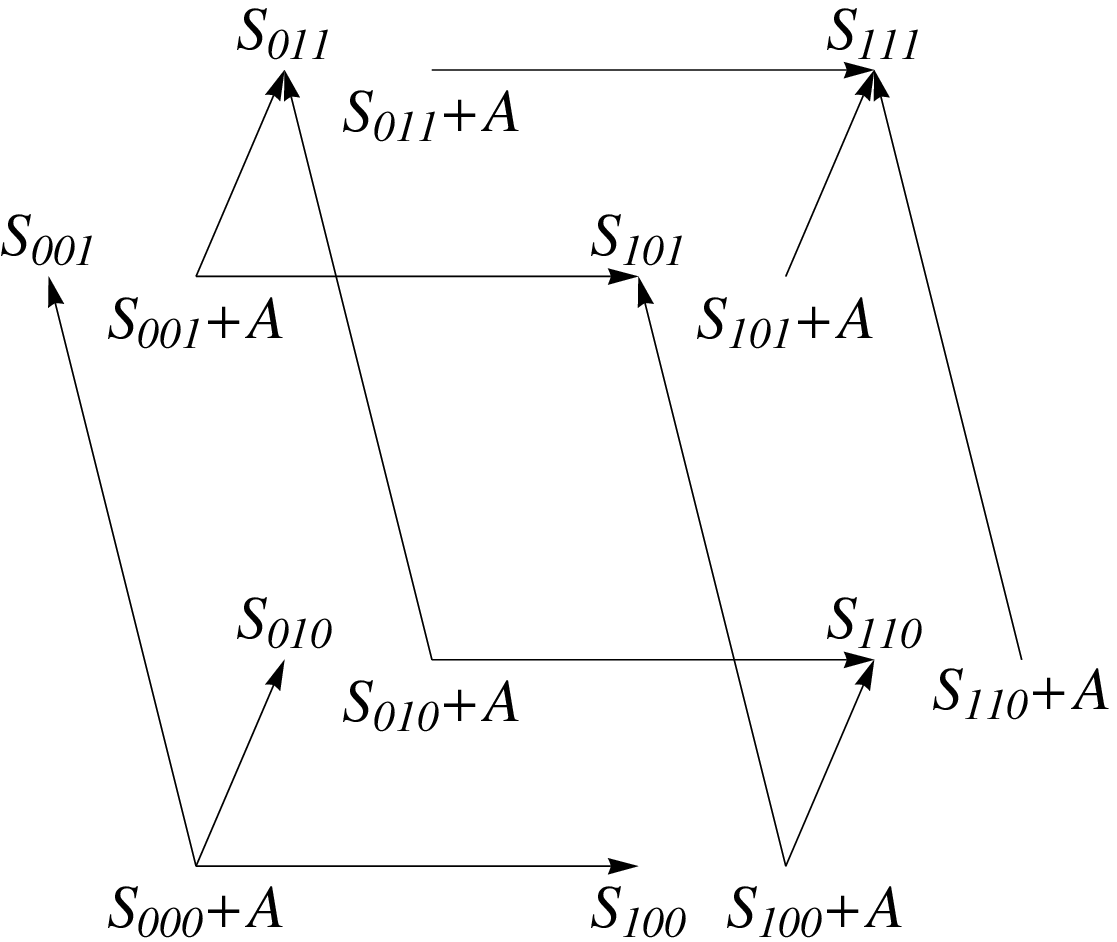}}
  \caption{}\label{fig:system4}
\end{figure}

\end{Exa}
\subsection{On the number of free parameters}
We would also like to make some comments on one of the statements in
Appendix 2 of \cite{colquhoun1}.
According to this, the
number of free parameters can be determined as follows.
Suppose that
we have a system with
\begin{itemize}
\item $N$ complexes,
\item $R$  rate coefficients (as parameters),
\item and $C$ constraints (the sum of the number of the microscopic
reversibility constraints and the number of arbitrary constraints---independent
of the microscopic
reversibility constraints and of each other---to be imposed on some
of the rate coefficients).
\end{itemize}
The number of free parameters will then be equal
to $R-\varrho,$ where $\varrho$ is the rank of an $C \times N$ matrix, $A$.

Recall from \cite{feinberg} that a reversible mass action system is
detailed balanced if and only if the rate constants satisfy the
$P-(N-L)$ circuit conditions and the $\delta$ spanning forest
conditions where the system has $P$ reaction pairs, $N$ complexes,
$L$ linkage classes, and $S$ is the rank of the stoichiometric
space, and $\delta$ is the deficiency of the network. Using these notations,
it can be written that the number of unknowns $R$ equals $2P$, and the number
of independent constraints $C$ equals
\begin{equation}
Q+(P-(N-L))+\delta=Q+P-S, \label{19}
\end{equation}
where $Q$ denotes the number of (further independent) external
constraints to be imposed on some of the rate coefficients. In
\cite{feinberg}, only $P-(N-L))+\delta=P-S$ is considered to be the
number of constraints and in \cite{colquhoun1}, the deficiency is
not taken into account in this sum.
Thus, our equation \eqref{19} is a
common generalization of the equations by Feinberg and Colquhoun et
al.

\section{Discussion, open problems}
We have provided a method to transform the most common ion channel
models into a model where mass-conservation is taken into account.
Using the theorem by Feinberg we have also shown that the heuristic
method happens to lead to the same results, in spite of the fact
that it is based on imprecise assumptions.

All the original models in question have a rectangular grid structure
with zero deficiency, and all the transformed models have a deficiency equal to the
number of independent circuits in the original model. To put it another way,
the sum of deficiency and the number of independent circuits is invariant
under our transformation. The natural question arises if the same consequences
can be drawn with nonzero deficiency (and nonzero number of independent circuits, respectively) and
what can be said about reactions having an FHJ-graph of different structure.

\section{Acknowledgements}
The present work has partially been supported by the
European Science Foundation Research Networking Programme:
Functional dynamics in Complex Chemical and Biological Systems,
and also by the Hungarian National Scientific Foundation, No. 84060.
This work is connected to the scientific program of the "Development of quality-oriented and harmonized R+D+I strategy and functional model at BME" project. This project is supported by the New Sz\'echenyi Plan (Project ID: T\'AMOP-4.2.1/B-09/1/KMR-2010-0002).

Prof. P. \'Erdi has proposed us to approach the problem in the present paper
with the tools of chemical reactor network theory,
and Prof. T. T\'oth was kind to draw our attention
to the important reference \cite{naundorf}.
Discussions with Mr. B. Kov\'acs and Ms. A. Szab\'o were really useful.

\newpage

\section{Appendix: Reactions of rectangular grid structure}
Let us consider a special class of reversible compartmental systems
with species constructed from $D\in\N$ different atoms, say,
$G^1,G^2,\dots,G^D$, sitting on a receptor which will be omitted as
it plays no rule in the calculations.
Let us represent the species
$G_{x_1}^1G_{x_2}^2\dots G_{x_D}^D$ by the vector
\((x_1,x_2,\dots,x_D)\in\N_0^D,\) and suppose (this is the
speciality of the system) that we only have the following reaction
steps \emph{in terms of the atomic representation of the species}:
\begin{eqnarray}\label{badmulti}
(x_1,x_2,\dots,x_D)\rightleftharpoons (x_1,x_2,\dots,x_{d-1},x_d+1,x_{d+1},\dots,x_D)\\
(0\le x_d\le p_d-1, p_d\in\N; d=1,2,\dots,D).\nonumber
\end{eqnarray}
This means that the Feinberg--Horn--Jackson graph (FHJ graph) of the
reaction is a rectangular grid in the first orthant with
$\prod_{d=1}^D (p_d+1)$ vertex.

Such kind of reactions are often used when modeling ion channels
see Fig. \ref{fig:system4}
or \cite{deyoungkeizer}.

Realizing that atoms are not conserved in the above reaction, we try
to improve it by constructing a model without this fault but
reflecting the same physical reality. In order to do so we have to
introduce $D$ new, single-atom species, $G^d\, (d=1,2,\dots,D)$ and the new
reaction steps
\begin{equation}\label{bettermulti}
\eb_d+(x_1,x_2,\dots,x_D)\rightleftharpoons
(x_1,x_2,\dots,x_d+1,\dots,x_D),
\end{equation}
where $\eb_d$ is the $d$th element of the standard base.

To test if a general reaction is detailed balanced or not one has to
write down \(\delta\) number of circuit conditions and \(K\) number
of spanning forest conditions in terms of the reaction rate
constants which form a set of necessary and sufficient conditions
together.

If we are interested in detailed balancing of the first reaction
\eqref{badmulti} we should rather transform it to
\eqref{bettermulti} and have only the spanning forest conditions.
The astonishing fact, however, is that for these special reactions
not only the number of conditions are the same, but the conditions
themselves, as well.

Let us use the following notations:

\begin{center}
\begin{tabular}{llll}
$N:$        & the number of complex vectors (the number of vertices)\\
$P:$        & the number of reaction pairs (the number of edges)\\
$L:$        & the number of linkage classes (the number of connected components)\\
$S:$        & the dimension of the stoichiometric space\\
            &(the number of independent reaction steps)\\
$\delta:$   & the deficiency\\
$K:$        & the number of independent circuits\\
\end{tabular}
\end{center}

To get some experience with this kind of systems we summarize the
essential characteristics of these systems in two and three
dimensions and then formulate and prove the general formula.

\begin{Sta}\label{sta}
If $D=2$ then the formulas for system \eqref{badmulti} (left
column) and system \eqref{bettermulti} (right column) are the
following:
\begin{center}
\begin{tabular}{llllll}
$N$     &$=$&$(p+1)(q+1)$    &$N'$      &$=$&$2(p+q)+3pq$\\
$L$     &$=$&$1$             &$L'$      &$=$&$p+q+pq$\\
$S$     &$=$&$N-1$           &$S'$      &$=$&$S$\\
$\delta$&$=$&$N-L-S=0$       &$\delta'$ &$=$&$N'-L'-S'=K$\\
$K$     &$=$&$P-(N-L)=pq$    &$K'$      &$=$&$P'-(N'-L')=0$\\
\end{tabular}
\end{center}
where $P=P'=p(q+1)+(p+1)q$.

If $D=3$ then the formulas are
\begin{center}
\begin{tabular}{llllll}
$N$     &$=$&$(p+1)(q+1)(r+1)$      &$N'$      &$=$&$2(p+q+r)+3(pq+pr+qr)+4pqr$\\
$L$     &$=$&$1$                    &$L'$      &$=$&$(p+q+r)+(pq+pr+qr)+pqr$\\
$S$     &$=$&$N-1$                  &$S'$      &$=$&$S$\\
$\delta$&$=$&$0$                    &$\delta'$ &$=$&$K$\\
$K$     &$=$&$pq+pr+qr+2pqr$        &$K'$      &$=$&$0$\\
\end{tabular}
\end{center}
where $P=P'=p(q+1)(r+1)+(p+1)q(r+1)+(p+1)(q+1)r$, see Fig. \ref{fig:system4a} and \ref{fig:system4b} as an illustration.
\end{Sta}

\begin{Thm}
\begin{enumerate}
\item[\nonumber]
\item
The essential characteristics of reactions \eqref{badmulti}
(with its FHJ-graph as a rectangular grid) and
\eqref{bettermulti} are as follows.
\begin{center}
\begin{tabular}{llllll}
$N$     &$=$&$\displaystyle\prod_{d=1}^D (p_d+1)$                           &$N'$      &$=$&$\displaystyle\sum_{k=1}^{D}(k+1)p_{d_1}p_{d_2}\dots p_{d_k}$\\
$L$     &$=$&$1$                                                            &$L'$      &$=$&$\displaystyle\sum_{k=1}^{D}p_{d_1}p_{d_2}\dots p_{d_k}$\\
$S$     &$=$&$N-1$                                                          &$S'$      &$=$&$S$\\
$\delta$&$=$&$0$                                                            &$\delta'$ &$=$&$K$\\
$K$     &$=$&$\displaystyle\sum_{k=2}^{D}(k-1)p_{d_1}p_{d_2}\dots p_{d_k}$  &$K'$      &$=$&$0$\\
\end{tabular}
\end{center}
where each sum is taken with the restrictions
$1\leq d_1< d_2<\dots<d_k\leq D$.
\item
The circuit conditions for reaction \eqref{badmulti} are exactly
the same as the spanning tree conditions for reaction \eqref{bettermulti}.
\end{enumerate}
\end{Thm}
\Proof{In both systems the number of edges can be calculated as
\begin{eqnarray*}
P=P'&=&p_1(p_2+1)\dots(p_D+1)+(p_1+1)p_2(p_3+1)\dots(p_D+1)+\dots
+\\&&+(p_1+1)\dots(p_{D-1}+1)p_D =\\
&=&\sum_{k=1}^Dk p_{d_1}p_{d_2}\dots p_{d_k} \quad (1\leq d_1<
d_2<\dots<d_k\leq D)
\end{eqnarray*}
The number of independent circuits in a graph can be calculated as
$K=P-(N-L)$. Thus, using that $N=1+L'$, we obtain the formula for
$K$:
\begin{eqnarray*}
K&=&P-(N-L)=P-N+1=P-L'\\
&=&\sum_{k=1}^Dk p_{d_1}p_{d_2}\dots
p_{d_k}-\sum_{k=1}^{D}p_{d_1}p_{d_2}\dots p_{d_k}
\end{eqnarray*}
The formulas for $N'$ and $L'$ follow from the following
observation: in the graph of the transformed system the number of
components consisting of one edge (and two vertices) is
$p_1+p_2+\dots+p_D$; the number of components consisting of two
edges (and three vertices) is $p_1p_2+p_1p_3+\dots p_{D-1}p_D$;
etc.; the number of components consisting of $D$ edges (and $D+1$
vertices) is $p_1p_2\dots p_D$.
The equality $S=S'$ and the equivalence of the circuit conditions and spanning forest conditions follow
from the $D$ dimensional version of the lemma. Finally, using that $S'=S=N-1=L'$, $\delta'=N'-L'-S'$ and $K'=P'-(N'-L')$,
we obtain the formulas for $\delta'$ and $K'$.
}

\nocite{*}
\bibliography{channels}
\bibliographystyle{plain}
\end{document}